\documentclass[twocolumn,noshowpacs,preprintnumbers,amsmath,amssymb]{revtex4}
\usepackage{amsfonts}
\usepackage{amsmath}
\usepackage{graphicx}
\def \ee{\end{equation}}
\def \be{\begin{equation}}
\def \bea{\begin{eqnarray}}
\def \eea{\end{eqnarray}}
\def\lsim{\lower.5ex\hbox{$\; \buildrel < \over \sim \;$}}
\def\gsim{\lower.5ex\hbox{$\; \buildrel > \over \sim \;$}}
\begin{document}

\title{An Angular Formalism for Spin One Half.}

\author{Benjamin Koch and Nicol\'as Rojas}
\affiliation{Pontificia Universidad Cat\'olica de Chile.\\ Av. Vicu\~na Mackenna
4860. Macul.\\ Santiago de Chile.}

\date{\today}
\email{bkoch@fis.puc.cl, nrojas1@uc.cl}

\begin{abstract}
Understanding spin one half is a crucial 
issue in the De Broglie Bohm framework. 
In this paper a concrete relativistic realization
of spin one half in terms of angular coordinates is developed.
A Lagrange formulation is found, equations of motion are
derived, and Lorentz invariance is discussed.
\end{abstract}
%

\maketitle
\section{Introduction}
\subsection{General Picture}
The de Broglie Bohm interpretation of quantum mechanics (dBB) is an
attempt to describe quantum phenomena
in a deterministic way.
It allows to assign physical reality to
a system, even at the instances between 
two measurements.
In this sense it goes beyond the Copenhagen
interpretation of quantum mechanics.
In such a framework the description of physics is given in terms of 
{\it hidden variables}, which are physical quantities
that are not present in the usual interpretation of quantum mechanics. 
A general introduction to the approach
and a discussion of most of the conceptual issues can
be found in \cite{Holland1,Nikolic:2006az}.
This formulation has recently attracted
some interest due to its potential connection
to quantum gravity and unification schemes 
\cite{Santamato:1984qe,Shojai:2000us,Carroll:2004hs,Koch:2008hn,Abraham:2008yr,Koch:2010bz,Falciano:2010zz}.
However, the description of spin one half within this
framework proved to be hard and not unique 
\cite{CufaroPetroni:1984vw,CufaroPetroni:1984wn,Holland:1988et,Nikolic:2003yu,Nikolic:2007ih,Wetterich:2010eh}.
This article is based on the combination
of an angular concept for spin one half, 
with a special formulation of the Dirac equation.
Those two starting points are introduced in
the subsections (\ref{sec_Angles} and \ref{sec_Brown}).

\subsection{Angular representation of Pauli matrices}
\label{sec_Angles}

In \cite{Holland1,Holland:2008cc} a description
of non-relativistic spin $1/2$ in terms
of the Euler-angles of a rigid rotator is given.
The algebraic definitions of this formulation
are the basis of our work and thus they will be
shortly defined in this section.
In this formulation the role of
two component spinors is taken by scalar
functions with additional angular degrees
of freedom.
The key step is to introduce angular operators $\hat M^k$,
that act on those functions in such a way
that they reproduce the algebra of the
Pauli spin matrices $\sigma^k$
\be
\begin{array}{ccc}\label{analogy}
\left. \begin{array}{c}
\psi_{a}(x)\\
\mbox{spinor with}\\
\mbox{two components}\\
\hline
\sigma^k\\
\mbox{Pauli matrices}
\end{array}
\right\}
&\leftrightarrow &
\left\{ \begin{array}{c}
\psi(x,\alpha_b)\\
\mbox{scalar function with}\\
\mbox{angular dependence}\\
\hline
 2 \hat M^k \\
\mbox{angular operator}
\end{array}
\right.
\end{array}
\ee
The angles $\alpha_b=\alpha, \beta, \gamma$
are defined with respect to the external space axes and
the operators $\hat M_k$ are 
\be\label{lambda}
\begin{array}{ccl}
\hat M_1=&i(\cos \beta \partial_\alpha-\sin \beta \cot \alpha
\partial_\beta
+\sin \beta \,/\sin(\alpha) \partial_\gamma) \\
\hat M_2=&i(-\sin\beta \partial_\alpha-\cos \beta
\cot \alpha \partial_\beta+\cos \beta \,/\sin(\alpha) \partial_\gamma)\\
\hat M_3=&i\partial_\beta
\end{array}
\ee
\be\label{lambdaprime}
\begin{array}{ccl}
\hat M'_1=&i(\cos \gamma \partial_\alpha+\sin \beta \csc\,
\alpha \partial_\beta
-\sin \gamma \cot(\alpha) \partial_\gamma) \\
\hat M'_2=&i(\sin\gamma \partial_\alpha-\cos \gamma
\csc\, \alpha \partial_\beta+\cos \gamma \cot(\alpha) \partial_\gamma)\\
\hat M'_3=&i\partial_\gamma
\end{array}
\ee

Which reads in a matrix notation
\be\label{lambda1}
\hat M_k=i\mathbb{A}_{k a}\partial_{a}\quad.
\ee
\be\label{lambda1prime}
\hat M'_k=i\mathbb{A}'_{k a}\partial_{a}\quad.
\ee
Those operator-matrices obey the commutation relations
\be
[\mathbb{A}_{la}\partial_{a},\mathbb{A}_{mb}\partial_{b}]
=\epsilon^{klm}\mathbb{A}_{kc}\partial_{c}\quad,
\ee
\be
[\mathbb{A}'_{la}\partial_{a},\mathbb{A}'_{mb}\partial_{b}]
=-\epsilon^{klm}\mathbb{A}'_{kc}\partial_{c}\quad.
\ee
From those relations one sees 
that the operators $\hat M$ follow
the same commutation relations as $\sigma_k/2$
\be\label{komm}
[\hat M_i,\hat M_j]=i\epsilon^{ijk}\hat M_k\quad,
\ee
while the operators $\hat M'$ follow 
anomalous commutation relations
\be
[\hat M'_i,\hat M'_j]=-i\epsilon^{ijk}\hat M'_k\quad. 
\ee
At this point it is interesting to note that
those operators do not fulfill the anti-commutation
relations in general. This only happens when
the operators are applied to functions with
special properties.
Holland finds the following complete set
of eigenfunctions
for the operator $\hat M^2$ with the eigenvalue
$s (s+1)=3/4$
\be\label{eigenfkt}
u_i(\alpha_b)=\left\{
\begin{array}{ccl}
 u_+&=&\frac{1}{\pi 2\sqrt{2 }}\cos (\alpha/2)e^{-i(\gamma+\beta)/2}\\
 u_-&=&\frac{-i}{ \pi 2\sqrt{2}}\sin (\alpha/2)e^{i(-\gamma+\beta)/2}\\
 v_+&=&\frac{-i}{ \pi 2\sqrt{2}}\sin (\alpha/2)e^{i(\gamma-\beta)/2}\\
 v_-&=&\frac{1}{ \pi 2\sqrt{2 }}\cos (\alpha/2)e^{i(\gamma+\beta)/2}\\
\end{array}
\right.
\ee
These functions differ by their eigenvalues 
$\pm 1/2$ with respect
to both, $\hat M_3$ and $\hat M'_3$.
Now, when acting on this set of functions, the operators $\hat M$
fulfill the Clifford algebra of the Pauli matrices $\sigma_k/2$
\be\label{antik}
\{\mathbb{A}_{ka}\partial_{a},\mathbb{A}_{lb}\partial_{b}\}u_i
+\frac{1}{2}\delta_{kl}u_i=0\quad,
\ee
\be
\{\mathbb{A}'_{ka}\partial_{a},\mathbb{A}'_{lb}\partial_{b}\}u_i
+\frac{1}{2}\delta_{kl}u_i=0\quad.
\ee
This technique is typically only applied
to non-relativistic spinors with two components 
\cite{Holland1,Holland:2008cc}.
But as it will be shown in the following sections it is also
sufficient for a description of relativistic
two component spinors.

\subsection{Lagrangian Formulation for Relativistic Spin 1/2}
\label{sec_Brown}
In order to have a direct applicability of the 
concepts in the previous section we seek a
quadratic and two component formulation of spin
one half. Such a formulation was given in 1958 by 
Laurie Brown \cite{Brown:1958zz},
who found a two component fermion theory 
with a Lagrangian 
which seemingly can not be found for the a single
Feynman-Gell-Mann equation \cite{Feynman:1958ty,CufaroPetroni:1985tu}.
One defines
\be
p^+=p_0+\vec{\sigma}\vec{p}\equiv\bar{\sigma}^\mu p_\mu\quad,
\ee
\be
p^-=p_0-\vec{\sigma}\vec{p}\equiv\sigma^\mu p_\mu\quad.
\ee
Due to the Clifford algebra of the Pauli matrices one has
\be
\bar{\sigma}_\mu \sigma_\nu = g_{\mu \nu}+h_{\mu \nu}
\ee
\be
\sigma_\mu \bar{\sigma}_\nu = g_{\mu \nu}+h'_{\mu \nu}\nonumber
\ee
with
\be
h_{0i}=-h_{i0} = -h'_{0i}=h'_{i0}=\sigma_i
\ee
\be
h_{ij}=h'_{ij} = -\frac{1}{2}[\sigma_i,\sigma_j]\nonumber
\ee
Further one finds from $D_\mu=p_\mu-A_\mu$
\be\label{Pipm}
D^+D^-=D^\mu D_\mu -\frac{i}{2}h_{\mu \nu}F^{\mu \nu}=
D^2-i\vec{\sigma}(\vec{E}+i\vec{B})
\ee
and
\be\label{Pipm2}
D^-D^+=D^\mu D_\mu -\frac{i}{2}h'_{\mu \nu}F^{\mu \nu}\quad.
\ee
A Lagrangian is defined by introducing an auxiliary two component
spinor $\Omega$
\be\label{LBrown}
{\mathcal{L}}_{LB}=m^{-1}(D^+\Omega)^{\dag}(D^-\psi)-m \Omega^\dag\psi
+ c.c.
\quad,
\ee
After a partial integration this Lagrangian can
be written in the form
\be\label{LDir}
{\mathcal{L}}_{LB}=
 (D_\mu \Omega)^\dag D^\mu \psi-\frac{i}{2}e\Omega^\dag 
 H^{1/2}_{\mu \nu}
F^{\mu \nu}\psi -m^2\Omega^\dag \psi+c.c.\;,
\ee
where
\be
H^{1/2}_{0i}=-H^{1/2}_{i0}=\sigma_i\,,\quad
H^{1/2}_{ij}=-\frac{1}{2}[\sigma_i,\sigma_j]\quad.
\ee

By varying with respect to the bi-spinorfields one obtains two 
 equations of motion which are apart from one term
the Feynman-Gell-Mann equations
\be\label{eomLB1}
m^2\psi=D^+D^-\psi=(D^\mu D_\mu -\frac{i}{2}h_{\mu \nu}
(F^{\mu \nu})\psi
\ee
\be\label{eomLB2}
m^2\Omega=D^+D^-\Omega=(D^\mu D_\mu -\frac{i}{2}h'_{\mu \nu}
F^{\mu\nu})\Omega\,.
\ee
The equations contain positive and
negative energy solutions. One can project
onto the positive solutions by
putting \cite{Brown:1958zz} 
\be\label{WeylEq}
iD^+\Omega=m\psi,\quad iD^-\psi=m\Omega \quad.
\ee
Under the discrete symmetries
C and P the equations (\ref{eomLB1}) and (\ref{eomLB2})
transform into each other (where C additionally has $\vec E \rightarrow -\vec
E$).
Those are the properties of the left handed and
right handed Weyl spinors.

The two equations in (\ref{WeylEq})
can be combined to a single equation for
a four component spinor
\be
(D_0-\tilde \alpha_i D_i-\beta m)
\left(\begin{array}{c}
 \Omega\\
\psi
\end{array}\right)=0 \quad,
\ee
with
\be
\tilde \alpha_i=
\left(
\begin{array}{cc}
 -\sigma_i & 0\\
0 & \sigma_i
\end{array}
\right),\quad
\beta=
\left(
\begin{array}{cc}
 0 & \it{1} \\
\it{1} & 0
\end{array}
\right)\quad.
\ee
By multiplying this with $\beta$
from the left hand side one obtains
\be\label{Dirac}
(\gamma^\mu D_\mu -m)
\left(\begin{array}{c}
 \Omega\\
\psi
\end{array}\right)=0
\ee
where $\gamma_\mu=(\beta,\beta \tilde\alpha_i)$.
Equation (\ref{Dirac}) is the Dirac equation
in Weyl representation.
This shows that at the level of equations of motion,
the action (\ref{LBrown})
is equivalent to Dirac's formulations of spin one half.

\section{An action for angles and spin 1/2}
In this section it will investigated whether
the formulation that was given in subsection \ref{sec_Angles}
can be directly and uniquely adapted to the
formulation of spin $1/2$ that was given in 
subsection \ref{sec_Brown}.
As first approach we propose the following action
\be
\label{LBrown2}
{\mathcal{S}}=\int d^4x \int d\omega\left({\mathcal{L}}_0+\Lambda^{1/2}
\right)+c.c\quad,
\ee
where $d\omega=\sin (\alpha)d\alpha d\beta d\gamma$
and the angular integrals are over 
$[0\le\alpha\le \pi,\, 0\le\beta\le 2\pi ,\, 0\le\gamma\le 4\pi]$.
The integral contains a main part
\be\label{L0}
{\mathcal{L}}_0=
m^{-1}(\hat{D}^+\Omega)^{\dag}(\hat{D}^-\psi)
-m \Omega^\dag\psi\quad,
\ee
the Lagrange multiplier
\begin{eqnarray}\label{Lmulti}
\Lambda^{1/2}=\lambda \cdot \Omega^\dag(\hat
M^2-\hbar^2\frac{3}{2})\psi \quad,
\end{eqnarray}
and the operators
\be\label{replace}
\hat D^{\pm} = D^\pm |_{\left[\sigma_k \rightarrow 2 i
\mathbb{A}_{ka}\partial_{a},\;\sigma_0={\it{1}}\rightarrow 1\right]}
\quad.
\ee
This is a straight forward guess, 
motivated by the form of the spinor action (\ref{LBrown}).
The two 
Lagrange multipliers ensure that the functions
$\psi$ and $\Omega$ are eigenfunctions of the
total spin operator $\hat M^2$ with the eigenvalue $s (s+1) =3/4$.
Having this constraint, the most general wave functions
can be expressed as linear combinations
of the $\hat M^2$ eigenfunctions given in (\ref{eigenfkt}).
With this representations of the functions $\psi$, $\Omega$
checks the following relations:
\be\label{orto}
\int d\omega\; u_i^* u_j=\delta_{ij}
\ee
One further finds two sets of Pauli matrices
\be\label{sig}
2 i \int d\omega\; u_i^*\mathbb{A}_{ka} \partial_{a}  u_j=
\left(\begin{array}{cc}
 \sigma_k &0\\
0& \sigma_k
\end{array}\right)_{ij}\quad,
\ee
one for $u_\pm$ and one for $v_\pm$.
Integrating this relation by parts
and using
\be\label{sig0}
2 i \int d\omega\; u_i^* (\partial_{a} \mathbb{A}_{ka}) u_j=0
\ee
one finds that 
\be\label{sigOm}
2 i \int d\omega\; (\partial_{a} u_i^*) \mathbb{A}_{ka}
  u_j=
-\left(\begin{array}{cc}
 \sigma_k &0\\
0& \sigma_k
\end{array}\right)_{ij}\quad.
\ee%
It is possible to write down many different
types of integrals, containing two
derivative operators $\partial_{a}$.
A selection of such integrals is 
The explicit results for those integrals are given
in the appendix \ref{AppendIntegrals}.
The remaining integrals can be obtained 
from (\ref{integrals}) by complex conjugation 
or partial integration.

A general spin one half expansion in terms of $M^2$ eigenfunctions 
is given by
\be\label{psiExpansion}
\psi(x,\alpha)=\sum_{a=1}^4 \psi_a(x)u_a(\alpha_b)\equiv
\psi_a(x) u^a(\alpha_b)\quad.
\ee
For the case of spin half it
is sufficient to stick to the eigenfunctions
of $u_+, u_-$ 
\be\label{wfkthalf}
\psi(x,\alpha_b)=\sum_{a=1}^2 \psi_a(x)u_a(\alpha_b)\quad.
\ee
Having the relations (\ref{orto}-\ref{ZeroForFerm1}) together
with (\ref{komm}) and (\ref{antik}) allows
to show that for the spin one half functions (\ref{wfkthalf})
\be\label{bigresult}
\int d\omega {\mathcal{L}}_0|_{s=1/2}=
{\mathcal{L}}_{LB}\quad.
\ee
This means that after integration over the angular
coordinates,
the Lagrangian (\ref{LBrown2})
with the replacements (\ref{replace})
is exactly identical to the Lagrangian (\ref{LBrown}).
After this integration the expansion coefficients
$\psi_a(x)$ from (\ref{wfkthalf}) become
the spinor components in (\ref{LBrown}).
The same result without any mixing 
terms is obtained if
the eigenfunctions $v_+, v_-$ are
used.

Thus, with the action 
(\ref{LBrown2}) an explicit Lagrange formulation of relativistic
spin one half in terms of the angles $\alpha$, $\beta$, and $\gamma$
has been found.
However, the identities
(\ref{antik} and \ref{ZeroForFerm1})
allow to write this angular Lagrangian
in many different forms.
Further ambiguities could in principle arise
when doing the replacement 
$\sigma^k\rightarrow 2 i\mathbb{A}_{ka}\partial_{a}$
within the action. It is a priori not clear whether
the derivative $\partial_{a}$ should act to the right, or to the left,
whether it should act on the functions $\psi$ and $\Omega$ only, 
or whether derivatives of the kind 
$\partial_{a}\mathbb{A}_{kb}$
should also be allowed.
A careful analysis of the integrals
(\ref{sig}-\ref{integrals}) shows that defining the
derivatives as ones that always act to the right 
gives the right result. When using other definitions,
one has to take possible sign changes like in
(\ref{sig} and \ref{sigOm}) into account.

\section{The equations of motion}
The equations of motion for the spinors
$\psi$ and $\Omega$ were given in (\ref{eomLB1}) and (\ref{eomLB2}).
Now we will turn to the angular dependent
Lagrangian (\ref{LBrown2}). Using the Clifford algebra (\ref{antik}), 
the main part of (\ref{LBrown2}) can be written in the form
\begin{eqnarray}
\mathcal{L}_0 &=& m^{-1}\left((D_\mu
\Omega)^{\dagger}D^{\mu}\psi - 2i\Omega^{\dagger}
F_{\mu \nu}{\mathbb{C}^{\mu\nu}}_k\hat{M}^k \psi \right)\nonumber \\
 &-& m\Omega^{\dagger}\psi\label{SlagB}
\end{eqnarray}
where $\mathbb{C}^{\mu\nu}_k$ is defined as
\be
\mathbb{C}_{00}^k=0,\,
\mathbb{C}_{0j}^k=-\mathbb{C}_{j0}^k=i\delta^k_j,\,
\mathbb{C}_{ij}^k=\epsilon^k_{ij} \quad.
\ee
%

%
We will use this Lagrangian in order to derive
angular equations of motion for spin one half.
When doing this one has to take into account that
the Lagrangian contains explicit dependence on angular degrees of freedom.
The Lagrangian further contains
 derivatives of order one in the angles. Note hat this Lagrangian
is only valid if the constraints (\ref{LBrown2}) are
applied, since only in this case the Clifford algebra (\ref{antik})
can be used. Using this, the equation of motion for the auxiliary field
$\Omega(x,\alpha_b)$ reads
\begin{eqnarray}\label{eomAux}
\frac{\partial \mathcal{L}}{\partial \Omega^{\dagger}} - \frac{D}{D x^{k}}\left( \frac{\partial \mathcal{L}}{\partial \Omega^{\dagger}_{,k}} \right) &=& 0 \quad,
\end{eqnarray}
where the total derivative as is given in the appendix (\ref{totDerApp}) is
defined as
\begin{eqnarray}\label{totDer}
 \frac{D}{D x^{k}}&=&\left( \frac{\partial}{\partial x^{k}} +
\frac{\partial \psi_a}{\partial x^{k}} \frac{\partial}{\partial \psi_a} + 
\frac{\partial \psi_{a,\sigma}}{\partial x^{k}} 
\frac{\partial}{\partial
\psi_{a,\sigma}}
 \right.\\
&&\quad \quad
\left.+ \frac{\partial \Omega_a}{\partial x^{k}} \frac{\partial}{\partial
\Omega_a} + \frac{\partial \Omega_{a,\sigma}}{\partial x^{k}} 
\frac{\partial}{\partial \Omega_{a,\sigma}}\right)\quad.\nonumber
\end{eqnarray}
For the case of (\ref{eomAux}) only the first line of (\ref{totDer})
contributes, while the second line of (\ref{totDer}) gives zero.
First, one evaluates derivatives of the Lagrangian:
\begin{eqnarray}
\frac{\partial \mathcal{L}}{\partial \Omega^{\dagger}} &=&  m^{-1}\left(-ieA_{\mu}D^{\mu}\psi - 2ie A_{\mu}\mathbb{C}^{\mu\nu}_k\hat{M}^{k}D_{\nu}\psi  \right) \nonumber \\
                                                       &&- m\psi \\
\frac{\partial \mathcal{L}}{\partial \Omega^{\dagger}_{,\rho}} &=& m^{-1}\left( D^{\rho}\psi - 2i\mathbb{C}^{\rho\nu}_k\hat{M}^{k}D_{\nu}\psi \right)\quad.
\end{eqnarray}
Expanding the fields according to (\ref{psiExpansion}),
this can be rewritten giving
\begin{eqnarray}
\frac{\partial \mathcal{L}}{\partial \Omega^{\dagger}} &=& 
m^{-1}\left(-ieA_{\mu}D^{\mu}\psi_{a}(x)u^a(\alpha_b) - \right. \nonumber \\
                                                       &-& \left.
2eA_{\mu}\mathbb{C}^{\mu\nu}_k\hat{M}^{k}u^a(\alpha_b)D_{\nu}\psi_{a}(x) 
\right) - \nonumber \\ 
&-&m\psi_{a}(x)u^a(\alpha_b)\label{LdAux} \\ 
\frac{\partial \mathcal{L}}{\partial \Omega^{\dagger}_{,\rho}} &=& m^{-1}\left(
D^{\rho}\psi_{a}(x)u^a(\alpha_b) - \right. \nonumber \\
                                                               &-&
\left.2i\mathbb{C}^{\rho\nu}_k\hat{M}^{k}u^a(\alpha_b)D_{\nu}\psi_{a}(x) \right)
\label{LdAuxd}
\end{eqnarray}

Joining (\ref{eomAux} - \ref{LdAuxd}) with the
corresponding terms from the Lagrange multipliers
one obtains the equation of motion from varying with
respect to $\Omega^\dagger$,
\begin{eqnarray}\label{eomprefinal}
D_{\rho}D^{*\rho}\psi(x_\mu,\alpha_b) -
e F_{\mu\nu}\mathbb{C}^{\mu\nu}_{k}\hat{M}^{k} \psi(x_\mu,\alpha_b)
\quad \quad\\ \nonumber
+ m^2\psi(x_\mu,\alpha_b) = 
m \lambda(\hat
M^2-\hbar^2\frac{3}{2})\psi(x_\mu,\alpha_b).
\end{eqnarray}
the corresponding constraint equation from $\lambda$ is
\be\label{constOm}
\Omega^\dagger(\hat
M^2-\hbar^2\frac{3}{2})\psi(x_\mu,\alpha_b)=0\quad.
\ee
Expanding the functions $\Omega,\, \psi$
in terms of a full set of $\hat M^2$ eigenfunctions
one finds that due to equation (\ref{constOm}) only
the functions (\ref{eigenfkt}) survive.
Thus the equation of motion (\ref{eomprefinal})
subject to the constraint (\ref{constOm}) reads
\begin{eqnarray}\label{eomfinal}
D_{\rho}D^{*\rho}\psi_{a}(x)u^a(\alpha_b) -
e F_{\mu\nu}\mathbb{C}^{\mu\nu}_{k}\hat{M}^{k}u_a(\alpha_b) \psi^a(x) 
\quad  \\ \nonumber
+ m^2\psi_{a}(x)u^a(\alpha_b) = 
0.
\end{eqnarray}
Similarly, the equation of motion for $\Omega$ can
be obtained by varying the action (\ref{SlagB}) with respect
to $\psi$ and the constraint $\lambda^*$. 
This calculation can be largely simplified
by performing a partial integration with respect
to all angular coordinates $\partial_{a}$ before
evaluating the equations of motion.
One finds that the analog equation of motion for $\Omega$
\begin{eqnarray}\label{eomfinal2}
D_{\rho}D^{*\rho}\Omega^{}_{a}(x)u^{a}(\alpha^*_b) -
e F_{\mu\nu}\mathbb{C}^{*\mu\nu}_{k}\hat{M}^{k}u^{}_a(\alpha^*_b)\Omega^{a}(x) 
\quad \quad \quad \\
+ m^2\Omega^{}_{a}(x)u^a(\alpha^*_b) = 
0\quad,
\nonumber
\end{eqnarray}
where $\mathbb{C}^{*\mu\nu}_k$ is the complex conjugation
of $\mathbb{C}^{\mu\nu}_k$. 
One confirms that the equations
(\ref{eomfinal},\ref{eomfinal2}) are exactly what one would
have obtained from the spinor equations of motion by the
replacement $\sigma^k \rightarrow 2\hat{M}^k$.
Since all
angles are real valued, the distinction
between angles $\alpha_b$ and complex conjugated
angles $\alpha_b^*$ is not important at this
point, it is however crucial in the
context of Lorentz invariance as it will be explained in the
following section.

\section{Lorentz invariance}
\subsection{Lorentz transformations 
as coordinate transformation in angular space}

Knowing that spinors in the standard picture
transform under a certain
representation of Lorentz group, one has to 
find the corresponding transformation in the
angular picture.
The spinors in the Lagrangian (\ref{LDir})
transform under a Lorentz transformation 
$x'_\mu=\Lambda_\mu^{\;\nu} x_\nu$
according to
\bea\label{TransPsi}
\psi'(x')&=S_\psi(\Lambda) \psi(x)=&e^{\frac{i}{2}( \theta_k-i\xi_k
)\sigma_k}\psi(x)\\
\label{TransOm}
\Omega'(x')&=S_\Omega (\Lambda)\psi(x)=&e^{\frac{i}{2}( \theta_k+i\xi_k
)\sigma_k}\Omega(x) \quad,
\eea
where $\theta_k$ correspond to rotations
and $\xi_k$ correspond to boosts and the 
$S_\psi(\Lambda), \, S_\Omega(\Lambda)$ are matrices in
the spinor space.
Following the scheme (\ref{analogy}) this suggests
in the angular formulation
\bea\label{TransPsi1}
\psi'(x',\alpha_b)&=&e^{i( \theta_k-i\xi_k
)\hat M_k}\psi(x,\alpha_b)\\
\label{TransOm1}
\Omega'(x',\alpha_b^*)&=&e^{i( \theta_k+i\xi_k
)\hat M_k^*}\Omega(x,\alpha_b^*) \quad.
\eea
One can expand the operators in (\ref{TransPsi1} and
\ref{TransOm1})
for infinitesimal transformations
\begin{eqnarray}\label{expandexp}
 e^{i(\theta - i\xi)_k \hat{M}_k } &=& N\approx 
1 - \theta_k A_{ka} \partial_a +i \xi_k A_{ka} \partial_a \label{M1},\\
 e^{i(\theta + i\xi)_k \hat{M}_k^*} &=&\tilde{N} \approx 1 - \theta_k
A_{ka}^* \partial_a^* - i \xi_k A_{ka}^* \partial_a^* \label{M2} \;.
\end{eqnarray}

In any case both $\psi(x,\alpha_b^*)$ and $(\Omega(x,\alpha^*_b))^*$
are just scalar functions and as such they can only transform due to their coordinates.
Thus, every non trivial transformation like the one for spinors
in (\ref{TransPsi}, \ref{TransOm}) has to originate from
a transformation of the angular coordinates
\begin{eqnarray}\label{genSpinor0}
\psi'(x',\alpha_b) &=& \psi(x,\alpha_b')\\ 
\Omega'(x',\alpha_b^*) &=& \Omega(x,\alpha_b^{*'})\quad.
\end{eqnarray}
Here $\alpha_b'=\alpha_b+\delta \alpha_b$ are the angular coordinates
that are transformed 
by the action of a certain Lorentz group element.

This implies a unique transformation of the 
additional angular coordinates
\begin{eqnarray}\label{genSpinor}
\psi'(\alpha_b) &=& \exp \left( i(\theta_k-i\xi_k)\hat{M}_k \right)\psi(\alpha_b)\nonumber \\
                &\approx& \psi(\alpha_b) - (\theta_k-i\xi_k)\mathbb{A}^k_a \partial_a \psi(\alpha_b) \\
                &\approx& \psi(\alpha'_b) = \psi(\alpha_b + \delta \alpha_b) \nonumber \\
             &\approx& \psi(\alpha_b) + \partial_a \psi(\alpha_b) \delta \alpha_a \Longrightarrow \nonumber
\\
\delta \alpha_b &=& -(\theta_k-i\xi_k)\mathbb{A}^k_a \partial_{a} \alpha_b\quad.
\label{deltaalpha}
\end{eqnarray}
Note that
according Lies theorem, it is enough to consider the infinitesimal
transformations, which lead after iterations to any finite transformation.
Following this logic one finds the finite transformations of the
angles
\begin{eqnarray}\label{anglefin}
\alpha'_b &=& \exp \left(-(\theta_k-i\xi_k)\mathbb{A}^k_a \partial_{a}\right) \alpha_b \label{angleLG}
\end{eqnarray}
where $\alpha_b$ are the angles and $\theta_k, \xi_k$ are the
parameters of the Lorentz transformation.
This expansion shows clearly that rotations
change the real values of the angles while 
boosts actually introduce an imaginary part to the angles.
Thus, after a boost the distinction between $\alpha_a$
and $\alpha_a^*$ becomes important. Those complex
parameters could cause problems, since it would 
be ambiguous to have within a part of the integral
functions and operators that depend on both $\alpha_a$
and $\alpha_a^*$. However, using the definition (\ref{TransOm})
where $\Omega=\Omega(x,\alpha^*_a)$,
one exactly avoids this type of problem.
With reference to the eigenfunctions (\ref{eigenfkt})
one sees that $(\Omega(x,\alpha^*_a))^*$ is effectively a function
of the $\alpha_a$ and not of the $\alpha_a^*$.
Noting that all parts of the action combine $\Omega^*$
and $\psi$ (or the complex conjugation of both) one finds that no such ambiguity exists.
This defines for a given Lorentz transformation
the transformed (possibly complex) angles.
It is important to note that the integral $\int d\omega$ over
those (possibly complex) angles is still effectively three dimensional,
since it is simply a transformed version of the three dimensional
real integral.

\begin{figure}[bht]
\centering
\includegraphics[width=7cm]{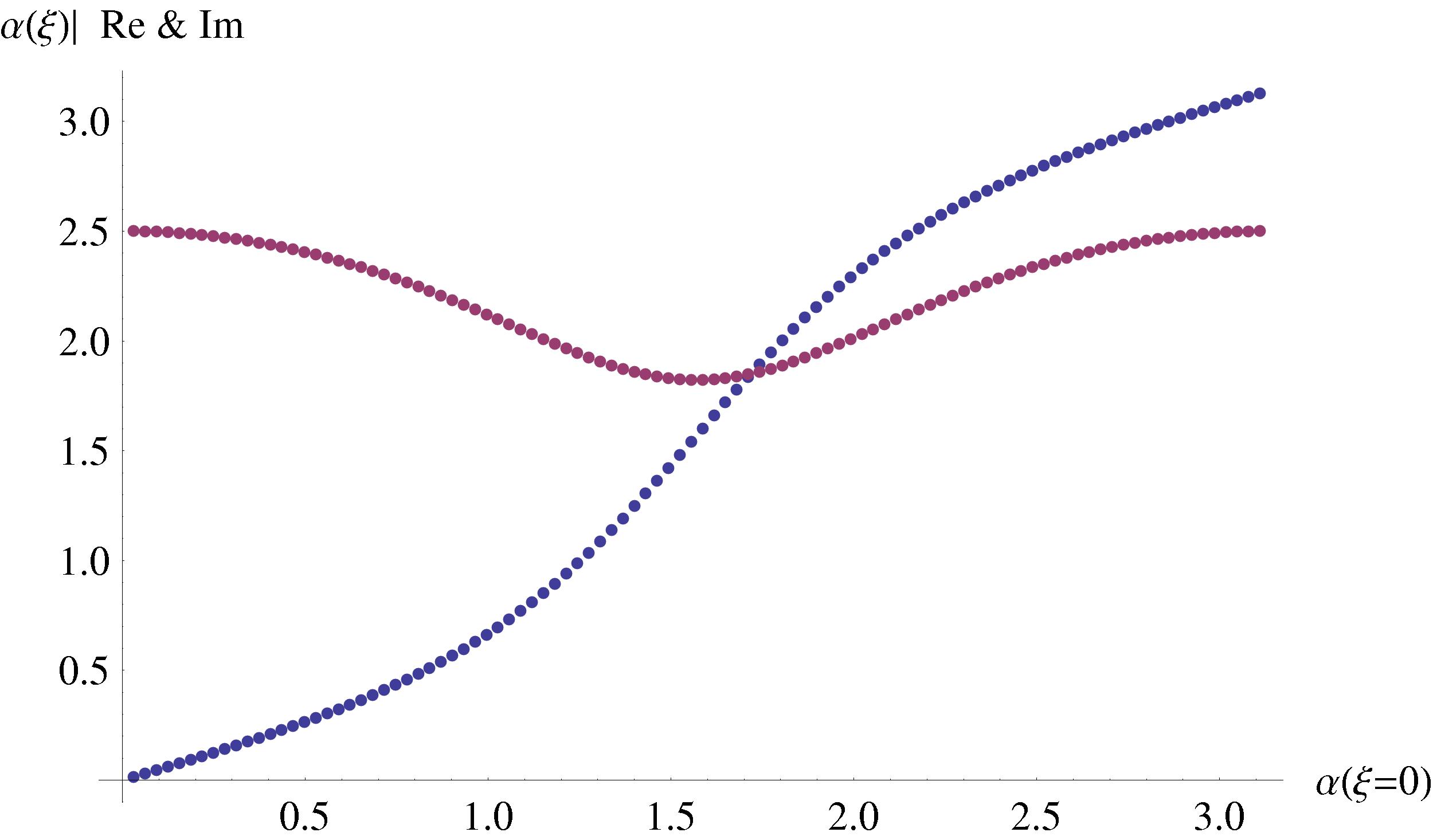}
\caption{Real (blue) and imaginary (red) part of the
angle $\alpha(\xi_1=2.5)$ boosted in x-direction
as a function of the original angle $\alpha(0): 0 \dots \pi$ 
for $\beta(0)=\pi/3$ and $\gamma(0)=0$.
}
\label{fig:alpha}
\end{figure}

\begin{figure}[bht]
\centering
\includegraphics[width=7cm]{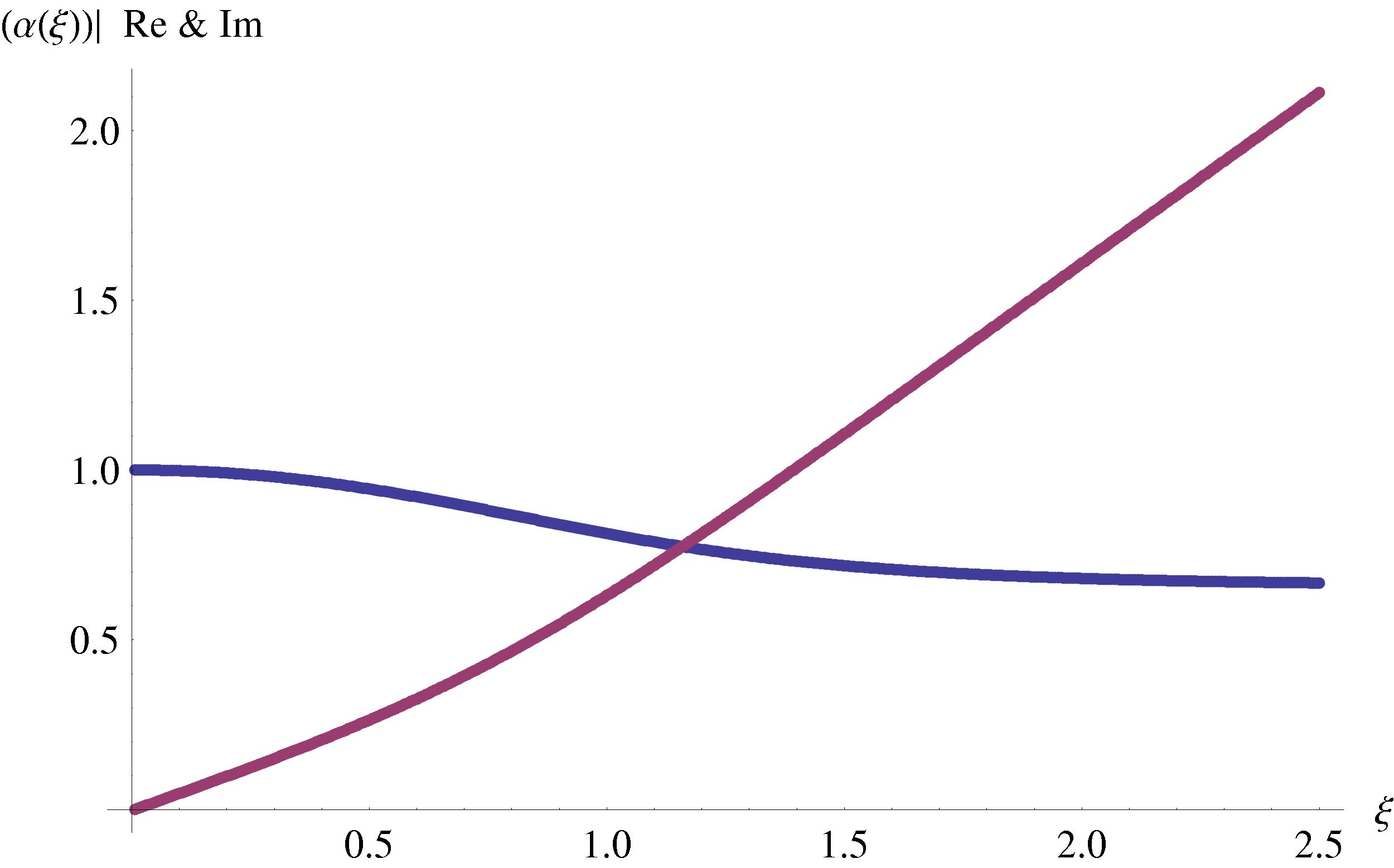}
\caption{Real (blue) and imaginary (red) part of the
angle $\alpha(\xi)$ boosted in x-direction
as a function of the boost factor $\xi: 0 \dots 2.5$.
The starting values were chosen to be
$\alpha(0)=1$, $\beta(0)=\pi/3$, and $\gamma(0)=0$.
}
\label{fig:alphaL}
\end{figure}

Please note that for rotations $\theta_k$ the angular transformation
behavior (\ref{angleLG}) is just an
application of a rotation in the Euler angles.
For boosts $\xi_k$ however the transformation (\ref{angleLG})
generates an imaginary part for the angles and it also
distorts their real part.
This behavior is exemplified in figure (\ref{fig:alpha})
where the real and the imaginary part of an angle
is plotted versus its value before the boost.
%
In figure (\ref{fig:alphaL}) it is shown
how those final values evolve as function of
the boost parameter $\xi$.
%
One observes that on the one hand the real part of the
internal parameters $\alpha_b$
changes as one would expect from a boosted coordinate
system. The imaginary part, on the other hand, reflects
a rescaling of the eigenfunctions (\ref{eigenfkt}) and their normality
conditions.

\subsection{Lorentz invariance of the angular Lagrangian}\label{sectlorinv}

From the expansion (\ref{expandexp}) one finds 
after a partial integration in the angular coordinate
that the mass term stays invariant
under such a infinitesimal transformation
\be\label{LImass}
\left<\Omega'^* \psi'\right>=
\left<\Omega^* \psi \right> +
{\mathcal{O}}(\theta^2_k,\xi_k^2,\theta_k \xi_k) \quad.
\ee
The same procedure works for the kinetic term of the
Lagrangian (\ref{SlagB})
\be\label{LIkin}
\left<(D_\mu\Omega)'^* D^\mu\psi'\right>=
\left<(D_\mu\Omega)^* D^\mu\psi \right> +
{\mathcal{O}}(\theta^2_k,\xi_k^2,\theta_k \xi_k) \quad,
\ee
where the vector and spinor transformations
cancel independently.
Before going to the interaction term,
it is useful to discuss the fundamental
correspondence between $\sigma^k$ and $\hat M^k(\alpha_a)$.
Given the fact that the matrices $\sigma^k$ do
not transform at all it seems puzzling that the operators
$\hat M^k(\alpha_a)$, who contain functions
of the angles, actually do transform due to the
transformation of those angles. 
This puzzle persists even after an integration
over $d\omega$.
The solution to this lies in the fact that in this picture the transformation
of spatial and angular coordinates are connected.
Thus, one can consider the possibility that the spatial
index $k$ of the operators also transforms, given
that $\hat M_k(\alpha_a)$ is not a constant matrix any more.
In that case an infinitesimal Lorentz transformation  
of this operator can be written as
\be
\hat M_k'(\alpha'_a)\approx
\hat M_k(\alpha_a)+ \tilde \omega_l^{\;k} M_k(\alpha_a)
+(\partial_b \hat M_k(\alpha_a)) \delta \alpha_b \quad.
\ee
On the right hand side, $\tilde \omega_l^{\;k}$
is an infinitesimal transformation of the external
index $k$ and the $\delta \alpha_b$ term corresponds
to the transformation of the angular parameter 
that was found in (\ref{deltaalpha}).
Allowing for this possibility one can find a
$\tilde \omega_l^{\;k}$ such that 
both transformation terms cancel and the angular operator
is actually constant under Lorentz transformation, just
as the usual Pauli matrices
\be\label{PauliTransf}
\hat M_k'(\alpha'_a)=
\hat M_k(\alpha_a)\quad.
\ee
This condition implies 
\be
\tilde \omega_l^{\;k}=\frac{i}{2}(\theta_n-i \vartheta_n)
\epsilon^{nkl}\quad.
\ee
Thus, using the above definition all operators
$\hat M_k(\alpha_a)$ can be treated as constants under Lorentz
transformations.
It is straight forward to show that the commutation
and anti-commutation relations (\ref{komm}) and (\ref{antik})
are not affected by this kind of index and or angular
transformation.
This identity
turns out to be useful for the interaction term,
where one further
has to make use
of the second order angular integrals 
$I^4_{kl},\; I^5_{kl}$ and $I^6_{kl}$.
One finds that the resulting terms cancel exactly
with the terms from the transformation of 
the external electromagnetic field
$F'^{\mu \nu}=\Lambda^\mu_\rho \Lambda^\nu_\sigma F^{\rho \sigma}$
\be\label{LIia}
\left< \Omega'^{*} F'_{\mu \nu}\mathbb{C}^{\mu\nu}_k\hat{M}'^k(\alpha_b')\psi' \right>
=\left< \Omega^{*} F_{\mu \nu}\mathbb{C}^{\mu\nu}_k\hat{M}^k \psi \right>
 + {\mathcal{O}}(\theta,\xi)^2\;.
\ee
In this calculation the infinitesimal transformation
$\Lambda^{\;\rho}_{\sigma} \approx \delta^{\;\rho}_{\sigma} + \omega^{\;\rho}_{\sigma}$
with
\begin{eqnarray}
\omega_{\mu}^{\;\nu} &=& 
\left(\begin{array}{cccc} 0 & \xi_x & 
\xi_y & \xi_z \\ \xi_x & 0 &
\theta_z & -\theta_y \\ \xi_y & -\theta_z & 
0 & \theta_x \\ \xi_z & \theta_y
& -\theta_x & 0 \end{array}\right) \quad,\label{omega}
\end{eqnarray}
was used. Please note that the same proof, term by term, can be done for the other version
of the Lagrangian that was previously defined in
(\ref{LBrown2}-\ref{replace}).
Thus, from (\ref{LImass}-\ref{LIia})
one finds that the angular Lagrangian is invariant
under the operational transformations (\ref{TransPsi1} and \ref{TransOm1})
after integrating out the angular degrees of freedom.
Those integrations involve a volume element of the 
angular degrees of freedom. Thus, it remains to be shown that
also this volume element 
$d\omega(\alpha,\beta,\gamma)=\sin \alpha{}\,d\alpha{}d\beta{}d\gamma$
is invariant under a  Lorentz transformation 
\begin{eqnarray}
d\omega(\alpha',\beta',\gamma') &=&  d\omega(\alpha,\beta,\gamma)\quad.
\label{relvol}
\end{eqnarray}
The above identity is best shown infinitesimally where 
$\alpha'\approx \alpha+\delta \alpha$ and
\be
d\omega(\alpha',\beta',\gamma')\approx \sin{(\alpha+\delta \alpha)}
\left| \frac{\partial \alpha'_a}{\partial \alpha_b} \right| 
d\alpha{}d\beta{}d\gamma\quad.\label{jacob0}
\ee
With the angular transformation (\ref{angleLG})
one finds to first order in $\eta_k=\theta_k-i\vartheta_k$ that the Jacobian is given by
\begin{eqnarray}
\left| \frac{\partial \alpha'_a}{\partial \alpha_b} \right| \approx
1 + \eta_1 \cot \alpha \cos \beta - 
\eta_2 \cot \alpha \sin \beta + 
\mathcal{O}(\eta^2) \,.
\label{jacob1}
\end{eqnarray}
For the same transformation the $\sin{(\alpha+\delta \alpha)}$ 
function reads
\begin{eqnarray}
\frac{\sin \alpha'}{\sin \alpha } \approx   1  - \eta_1 \cot \alpha \cos \beta 
                  + \eta_2 \cot \alpha \sin \beta + \mathcal{O}(\eta^2)  \,.
                  \label{sin}
\end{eqnarray}
Joining (\ref{jacob0}, \ref{jacob1}, and \ref{sin}) one sees that the
relation (\ref{relvol}) holds.

After proving the invariance statements
(\ref{LImass}, \ref{LIkin}, \ref{LIia} and \ref{relvol})
by brute force computation it is instructive
to reflect, on what they actually mean in terms
of the coordinate transformation $\alpha_b \rightarrow \alpha_b'$.
In this context the relations (\ref{LImass}) and (\ref{LIkin})
are actually trivial since no free index is
involved and any arbitrary substitution of the
angular coordinates would have left the integrals invariant.
What is unique about the
angular transformations (\ref{anglefin}) is
that they leave the volume element $d\omega$
invariant and that they compensate the space-time
transformation of $F_{\mu \nu}$ in (\ref{LIia}).
This compensation in (\ref{LIia}) is possible
due to the simultaneous transformation of the index $k$
and the angles $\alpha_b$, which allowed 
to use the identity (\ref{PauliTransf}).

Thus, in the context of Lorentz transformations
three things have been shown:
\begin{itemize}
\item[($i$)]{Event though the angles $\alpha_b$
are transformed, 
Pauli matrices and 
angular operators can be treated analogue in the
sense the identity (\ref{PauliTransf})}
\item[($ii$)]{The Lagrangian (\ref{LBrown})
is Lorentz invariant, in both formulations}
\item[($iii$)]{
The transformation laws for the left handed and right
handed spinors $\psi_a(x_\mu)$, $\Omega_a(x_\mu)$
correspond to the same possibly complex
transformation of the angular coordinates
$\alpha_b$ in the scalar functions
$\psi(x_\mu,\alpha_b)$ and $\Omega(x_\mu,\alpha^*_b)$
}
\end{itemize}
Further checks concerning Lorentz invariance
are performed in (\ref{AppendL1}).

\section{Conclusions}
In this paper we presented a new formulation of
relativistic spin one half.
The corresponding Lagrangian
has three additional angular coordinates 
in terms of the Euler angles
($\alpha$, $\beta$, and $\gamma$).
In the construction
scalar functions
with angular dependence play the role of spinors
and angular operators play the role of
Pauli matrices.

It is shown that, after integrating out
the angular degrees for freedom,
the bi-spinorial action of Brown is obtained
(\ref{bigresult}).
Since this bi-spinor 
action is known to be equivalent to the
Dirac equation (\ref{Dirac}), the main objective
of finding a new de Broglie Bohm
formulation equivalent to the Dirac equation is achieved.
Further, a equation of motion (\ref{eomfinal}) for
the extended space (containing space-time 
and angular degrees of freedom) is derived.
This equation of motion should allow to calculate
the Bohmian trajectories in the extended space.
Finally, Lorentz invariance is shown
and it is demonstrated how boosts and rotations
act on the scalar functions 
$\psi(x_\mu,\alpha_b)$ and $\Omega(x_\mu,\alpha_b^*) $.
The beauty of this formulation lies in
the fact that Lorentz transformations
for spinors simply correspond to a change 
of the parameters $\alpha_a$ as indicated in 
(\ref{anglefin}).  \\

Future investigation will be on
explicit solutions of the equation of motion (\ref{eomfinal})
and on the applicability of the method to
systems with spin $\neq 1/2$.

Many thanks to M.A. Diaz, P. Holland,
C. Valenzuela, and the HEP-PUC group for valuable 
hints and discussions. 
The work of B. K. was supported by CONICYT
project PBCTNRO PSD-73. The work of
N. R. was supported by CONICYT scholarship
\section{Appendix}
\subsection{List of angular integrals}
\label{AppendIntegrals}
This part of the appendix contains the explicit 
form of angular integrals with two derivatives 
\be
\int d\omega \; u_i^* (\partial_b A_{kb})u_j=0\;,
\ee
\begin{eqnarray}\label{integrals}
 (I^0_{kl})_{ij}=&\int d\omega\; (\mathbb{A}_{la}
(\partial_{a}u_i))^*\mathbb{A}_{kb}
\partial_{b}  u_j\\ \nonumber
 (I^1_{kl})_{ij}=&\int d\omega\; (\mathbb{A}_{ka} (\partial_{a}u_i))^*
(\partial_{b} \mathbb{A}_{lb})
 u_j\\ \nonumber
 (I^2_{kl})_{ij}=&\int d\omega\; (u_i)^* (\partial_{a} \mathbb{A}_{ka}) 
(\partial_{b} \mathbb{A}_{lb})  u_j\\\nonumber
 (I^3_{kl})_{ij}=&\int d\omega\; (u_i)^* \mathbb{A}_{ka}(\partial_{a} 
\partial_{b} \mathbb{A}_{lb})  u_j\\\nonumber
 (I^4_{kl})_{ij}=&\int d\omega\; (u_i)^* \mathbb{A}_{ka}
\mathbb{A}_{lb}\partial_{a} 
\partial_{b}  u_j\\\nonumber
 (I^5_{kl})_{ij}=&\int d\omega\; (u_i)^* (\partial_{a} \mathbb{A}_{ka}) 
 \mathbb{A}_{lb}\partial_{b}  u_j\\\nonumber
 (I^6_{kl})_{ij}=&\int d\omega\; (u_i)^*\mathbb{A}_{ka} 
 (\partial_{a} \mathbb{A}_{lb}) 
 \partial_{b}  u_j\\\nonumber
 (I^7_{kl})_{ij}=&\int d\omega\; (u_i)^* (\partial_{b} \mathbb{A}_{ka}) 
 (\partial_{a}\mathbb{A}_{lb}) u_j\\ \nonumber
 (I^8_{kl})_{ij}=&\int d\omega\;(\partial_{a}u_i)^* 
\mathbb{A}_{ka} \mathbb{A}_{lb}
\partial_{b}  u_j \quad.
\end{eqnarray}
Evaluating those integrals one finds.
\begin{eqnarray}
\left(I^0_{kl}\right)_{ij}=\frac{1}{4}\left(\begin{array}{cc}
\sigma_l \sigma_k& 0\\
0&\sigma_l \sigma_k
\end{array}\right),\\ \nonumber
\left(I^1_{kl}\right)_{ij}=0,\\
\left(I^2_{11}\right)_{ij}=
-\frac{\delta_{ij}}{2}\lim_{\epsilon\rightarrow0}
(\cos(\epsilon)+\log (\tan (\frac{\epsilon}{2}))),\\ \nonumber
\left(I^2_{22}\right)_{ij}=\left(I^2_{11}\right)_{ij},
\left(I^2_{\mbox{rest}}\right)_{ij}=0,\\
\left(I^3_{11}\right)_{ij}=
\left(I^3_{22}\right)_{ij}=
\frac{1}{2} \delta_{ij},\, \left(I^3_{\mbox{rest}}\right)_{ij}=0,\\
\left(I^4_{kk}\right)_{ij}=
-\frac{1}{4} \delta_{ij},
\left(I^4_{12}\right)_{ij}=\left(I^4_{21}\right)_{ij}=0\\ \nonumber
\left(I^4_{13}\right)_{ij}=
\left(I^4_{31}\right)_{ij}=
\frac{1}{4}\left(\begin{array}{cc}
\sigma_3 \sigma_1& 0\\
0&\sigma_3 \sigma_1
\end{array}\right),\\ \nonumber
\left(I^4_{23}\right)_{ij}=
\left(I^4_{32}\right)_{ij}=
\frac{1}{4}\left(\begin{array}{cc}
\sigma_3 \sigma_2& 0\\
0&\sigma_3 \sigma_2
\end{array}\right),\\
\left(I^5_{kl}\right)_{ij}=0,
\end{eqnarray}
\begin{eqnarray}
\left(I^6_{kk}\right)_{ij}=\left(I^6_{23}\right)_{ij}=
\left(I^6_{13}\right)_{ij}=0,\\ \nonumber
\left(I^6_{21}\right)_{ij}=
-\left(I^6_{12}\right)_{ij}=
\frac{I}{4}\left(\begin{array}{cc}
\sigma_3 & 0\\
0&\sigma_3
\end{array}\right),\\ \nonumber
\left(I^6_{32}\right)_{ij}=
-\frac{1}{2}\left(\begin{array}{cc}
\sigma_3 \sigma_2& 0\\
0&\sigma_3 \sigma_2
\end{array}\right),\\ \nonumber
\left(I^6_{31}\right)_{ij}=
-\frac{1}{2}\left(\begin{array}{cc}
\sigma_3 \sigma_1& 0\\
0&\sigma_3 \sigma_1
\end{array}\right),
\end{eqnarray}
\begin{eqnarray}
\left(I^7_{11}\right)_{ij}=
-\frac{\delta_{ij}}{2}\lim_{\epsilon\rightarrow0}
(\cos(\epsilon)+\log (\cot (\frac{\epsilon}{2}))),\\ \nonumber
\left(I^7_{22}\right)_{ij}=\left(I^7_{11}\right)_{ij},
\left(I^7_{\mbox{rest}}\right)_{ij}=0,\\
\left(I^8_{kl}\right)_{ij}=
\frac{1}{4}\left(\begin{array}{cc}
\sigma_k \sigma_l& 0\\
0&\sigma_k \sigma_l
\end{array}\right).
\end{eqnarray}
For the spin one half eigenfunctions one finds
further the useful relations
\be
\label{sigsig}
4(I^8_{kl})_{ij}=-4\left((I^4_{kl})_{ij}+(I^6_{kl})_{ij}\right)=
\left(\begin{array}{cc}
 \sigma_k \cdot \sigma_l&0\\
0& \sigma_k \cdot \sigma_l
\end{array}\right)_{ij}\quad,
\ee
\be\label{ZeroForFerm1}
(I^5_{kl})_{ij}=\left((I^2_{kl})_{ij}+(I^7_{kl})_{ij}\right)/2+(I^3_{kl})_{ij}=0
\quad.
\ee
When using partial integration one has to
take care, since in many cases the boundary
terms do not vanish.
The integrals $\left(I^2_{11}\right)_{ij}$
and
$\left(I^7_{11}\right)_{ij}$
do not converge at the boundaries of $\alpha$.
Therefore the asymptotic result for $\int_{0+\epsilon}^{D-\epsilon} d\alpha
\dots$ was given.

\subsection{EOM's from a Lagrangian with Explicit Coordinate Dependence}
%
In this appendix the equations of motion for a physical system 
that is defined in a region $R$ and whose
coordinates are $x_k$ ($k=1..n$) is discussed. 
The fields are functions of these
variables and they are labeled as $\psi^{\alpha}$, 
where $\alpha$, runs over all
fields defined in the system. 
The discussion follows basically  \cite{Hill:1951zz}.
The Lagrange density 
 depends explicitly 
on $x_k$, $\psi^{\alpha}$, 
and $\partial_k\psi^{\alpha}$.
The action is defined as:
\begin{eqnarray}
\mathbb{A} &=& \int d(x) \mathcal{L}\left[ x_k,  \psi^{\alpha}\left( x \right),
\partial_{k}\psi^{\alpha}\left( x \right) \right]\quad.
\end{eqnarray}

When performing a variation one has to take into account 
the region $R$ and the coordinates $x_k \rightarrow x_k + \delta x_k$.
Thus, in general a field if varied according to
\begin{eqnarray}
\psi^{'\alpha} ( x' ) &:=& \psi^{\alpha} 
(x+ \delta x) + \delta \psi^{\alpha} = \nonumber \\
           &=& \psi^{\alpha} (x) + \partial_{k}\psi^{\alpha} (x)
\delta x^{k} + \delta \psi^{\alpha} \Rightarrow \nonumber \\
\Rightarrow \delta \psi^{\alpha} (x) 
&:=& \psi^{'\alpha} ( x' ) - \psi^{\alpha}(x) = \nonumber \\
   &=& \partial_{k}\psi^{\alpha} (x) \delta
x^{k} + \delta \psi^{\alpha} \\
\Rightarrow \delta \partial_{k}\psi^{\alpha} (x) &=& \partial_{l}
\partial_{k}\psi^{\alpha} (x)\delta x^{l} + \delta
\partial_{k}\psi^{\alpha}\quad.
\end{eqnarray}

Performing the total variation in the action, one gets
\begin{eqnarray}
\delta \mathbb{A} &=& \int d(x + \delta x) \mathcal{L} \left[x_k + \delta x_k, \psi^{\alpha}(x) \right.
\nonumber \\
            &&+ \left. \delta \psi^{\alpha} (x) ,
\partial_{k}\psi^{\alpha} (x) + \delta \partial_{k}\psi^{\alpha} (x)\right] -
\nonumber \\
      &&- \int d(x) \mathcal{L}\left[ x_k,  \psi^{\alpha}\left( x
\right), \partial_{k}\psi^{\alpha}\left( x \right) \right]  \nonumber\\
\nonumber
&\Rightarrow&
\end{eqnarray}
\begin{eqnarray}
\delta \mathbb{A} &=& \int d(x)\\ \nonumber
&&\left( \frac{D}{D x_k}\left[ \mathcal{L}\delta x_k + \frac{\partial
\mathcal{L}}{\partial \psi^{\alpha}_{,k}} ( \delta \psi^{\alpha}(x) -
\psi^{\alpha}_{,k} (x) \delta x^{k} ) \right] \right. \nonumber \\ \nonumber
    &&+ \left. \left( \frac{\partial \mathcal{L}}{\partial
\psi^{\alpha}} - \frac{D}{D x_k}\frac{\partial \mathcal{L}}{\partial
\psi^{\alpha}_{,k}} \right)( \delta \psi^{\alpha}(x) - \psi^{\alpha}_{,k} (x)
\delta x^{k} )    \right). \label{actvar}
\end{eqnarray}
Here, the derivative $\frac{D}{D x_k}$ means the total derivative respect to the
coordinates
\begin{eqnarray}\label{totDerApp}
\frac{D}{D x_k} &=& \frac{\partial}{\partial x_k} +
\psi^{\alpha}_{,k}\frac{\partial }{\partial \psi^{\alpha}} + 
\psi^{\alpha}_{,kl}\frac{\partial }{\partial \psi^{\alpha}_{l}} \quad.
\end{eqnarray}
The first total derivative in (\ref{actvar}) is a boundary term, 
which can be neglected.
If one demands that the region $R$ is
left invariant one can set $\delta x_k = 0$.
The remaining variation of the action is then
\begin{eqnarray}
\delta \mathbb{A} &=& \int d(x)\left(  \left( \frac{\partial
\mathcal{L}}{\partial
\psi^{\alpha}} - \frac{D}{D x_k}\frac{\partial \mathcal{L}}{\partial
\psi^{\alpha}_{,k}} \right) \delta \psi^{\alpha}(x)    \right).
\end{eqnarray}
Hamilton's principle requires that the above expression
must vanish for every choice of the region of integration
and for every choice of the variational functions $\delta \psi^{\alpha}(x)$.
Thus the resulting equations of motion are 
\begin{eqnarray}
\frac{\partial \mathcal{L}}{\partial \psi^{\alpha}} - \frac{D}{D x_k}
\frac{\partial \mathcal{L}}{\partial \psi^{\alpha}_{,k}} &=& 0 \quad.\label{modEL}
\end{eqnarray}
The total derivative (\ref{totDerApp}) was defined in
such a way that
the equations of motion have a form that is
similar to the one that is normally used in physics. 
The equations here are, however,
more general since they include explicit
coordinate dependence.

\subsection{Lorentz group and its explicit form}
\label{AppendL1}

%
A covering group of 
the orthochronus Lorentz group $SL(2,\mathbb{C})$
is defined by $2\times 2$ matrices, with null trace
and determinant 1. This group is specified
by six real valued parameters.
As candidates to satisfy all this requirements, one may use:
\begin{eqnarray}
\left\{ L_k , K_k \right\} &\Leftrightarrow& \left\{ \frac{1}{2}\sigma_k ,
-\frac{i}{2}\sigma_k \right\} \nonumber \quad.
\end{eqnarray}

Which correspond to rotations ($L_k$) and boosts ($K_k$). 
They are the generators of Lorentz transformations in this space 
of representation of $\sigma$
matrices. The above implies that
an arbitrary element of $SL(2,\mathbb{C})$ can be written as
\begin{eqnarray}
\Lambda(\theta,\xi) &=& \exp i\left( \frac{1}{2}\theta_k\sigma_k +
-\frac{i}{2}\xi_k \sigma_k \right) \nonumber \\
                      &=& \exp \left(\frac{i}{2} \left( \theta_k -
i\xi_k\right)\sigma_k \right)\quad.
                      \label{trafoL}
\end{eqnarray}
This corresponds to one representation of the Lorentz group 
$SL(2,\mathbb{C})$, namely $(1/2,0)$. If one conjugates
the previous expression one obtains another representation, namely $(0,1/2)$,
which can be expressed as
\begin{eqnarray}\label{trafoLstar}
\Lambda^{*}(\theta,\xi) &=& \sigma_2 \left[ \exp \frac{i}{2} \left( \theta_k +
i\xi_k\right)\sigma_k \right] \sigma_2\quad.
\end{eqnarray}
Individually, both act in a two component spinor space (or a Weyl spinor). 
In this way, a vector in the 
representation space $(1/2,0)\otimes(0,1/2)$, transforms:
\begin{eqnarray}
\Psi'_D &=& \left( \begin{array}{cc} \Lambda (\alpha,\beta) & 0 \\ 0 & \sigma_2
\Lambda^{*}(\alpha,\beta) \sigma_2  \end{array} \right) \Psi_D
\end{eqnarray}
Those are Dirac spinors. For our purposes the formulation in 
\cite{Brown:1958zz}, is considering Weyl spinors, 
as it can be seen from (\ref{Dirac}).
This means that by choosing a representation for $\psi$, for
instance $(1/2,0)$, one immediately fixed the representation 
of $\Omega$ to be 
the other one ($(0,1/2)$). 
With this,  Lorentz 
invariance of the Lagrangian in subsection \ref{sectlorinv}
is preserved.
Remember that $\Omega$ is connected to $\psi$ via the equation
(\ref{WeylEq}).\\

A remaining question in our discussion
is whether the transformation
behavior given in (\ref{trafoL} and \ref{trafoLstar})
goes through to the angular formulation.
In order to see this
one may use the expectation values of the $\hat{M}$ spin operators in the
basis of eigenfunctions of $M_3$, $M_3'$ and $M^2$. 
This can be done since both $\Omega$ (living in $(0,1/2)$)
and $\psi$ (living in $(1/2,0)$) are expressed with the
same basis functions $u_a(\alpha)$. One has to show that the relation
\begin{eqnarray}\label{LImaster}
\left<u^*_i\right| 
e^{i(\theta - i\xi)_k \hat{M}^k}\left|u_j \right> &=& 
\left(e^{\frac{i}{2}(\theta-
i\xi)_k \sigma^k}\right)_{ij}\quad.
\end{eqnarray}
holds for those expectation values.
This proof involves excessive use of 
the Baker Campbell Hausdorff relation.
It will be given separately for pure rotations and pure boosts.
First, we consider $\xi_k = 0$:
\begin{eqnarray}
\left< e^{i\theta_k \hat{M}^k} \right>_{ab} &=& \left< \openone + \frac{1}{1!}
\left(i\theta_k\hat{M}^k\right) + \frac{1}{2!} \left(i\theta_k\hat{M}^k\right)^2
\right. \nonumber \\
                                            &+& \left.
\frac{1}{3!}\left(i\theta_k\hat{M}^k\right)^3 + \dots \right>_{ab}
\Longleftrightarrow \nonumber \\
\left< \openone \right>_{ab} &=& \delta_{ab} \\
\left< i\theta_k\hat{M}^k \right>_{ab} &=& \frac{1}{2}
i\theta_k(\sigma^k)_{ab} 
\end{eqnarray}
\begin{eqnarray}
\left< \left(i\theta_k\hat{M}^k\right)^2 \right>_{ab} &=& -\left<
\left(\theta_k\hat{M}^k\right)\left(\theta_l\hat{M}^l\right) \right>_{ab}
\nonumber \\
                                                 &=& -\frac{1}{2}\left<
\theta_k\theta_l([\hat{M}^k,\hat{M}^l] +  \{\hat{M}^k,\hat{M}^l\} ) \right>_{ab}
\nonumber \\
                                                 &=&
-\frac{1}{2}\left<\theta_k\theta_l\{\hat{M}^k,\hat{M}^l\}\right>_{ab} \nonumber
\\
                                                 &=& -\left( \frac{1}{2}
\right)^2\theta_k\theta^k \delta_{ab} \\
\left< \left(i\theta_k\hat{M}^k\right)^3 \right>_{ab} &=& -i\frac{1}{2}\left<
\theta_k\alpha_l\theta_{m}\hat{M}^k( [\hat{M}^l,\hat{M}^m] + \right. \nonumber
\\
                                                      &+& \left. 
\{\hat{M}^l,\hat{M}^m\} )\right>_{ab} \nonumber \\
                                                      &=&
-i\frac{1}{2}\left<\theta_k\theta_l\theta_{m}\hat{M}^k
\{\hat{M}^l,\hat{M}^m\}\right>_{ab} \nonumber
\end{eqnarray}
Applying the base of eigenfunctions one can replace the anti-commutator
by a Kronecker delta (see \ref{antik}).
This gives
\begin{eqnarray}
\left< \left(i\theta_k\hat{M}^k\right)^2 \right>_{ab} &=& -i\left( \frac{1}{2} \right)^3
\theta^2 \theta^k(\sigma_k)_{ab}\quad.
\end{eqnarray}
The remaining terms can be solved in the same way, with the same tricks. 
So, by induction, one can
also finish the entire table of expectation values 
of powers of $\hat{M}_k$ operators. Finally:
\begin{eqnarray}
\left< e^{i\theta_k \hat{M}^k} \right> &=&e^{\frac{i}{2}\theta_k \sigma^k} 
                      \quad. \label{exprot}
\end{eqnarray}
Using an identical procedure for pure boosts one finds
\begin{eqnarray}
\left< e^{\xi_k \hat{M}^k} \right> 
      &=& e^{\frac{1}{2}\xi_k \sigma^k}\quad.
\end{eqnarray}
Thus, it has been shown that (\ref{LImaster})
holds for rotations and boosts 
and that the Lorentz transformations
of Weyl spinors can be translated directly to 
the scalar functions with angular dependence.\\


\begin{thebibliography}{15}

\bibitem{Holland1} Peter Holland {\it The Quantum Theory of Motion}. Cambridge University Press. (1994)

\bibitem{Nikolic:2006az}
  H.~Nikolic,
  Found.\ Phys.\  {\bf 37}, 1563 (2007)
  [arXiv:quant-ph/0609163].

\bibitem{Santamato:1984qe}
  E.~Santamato,
  J.\ Math.\ Phys.\  {\bf 25}, 2477 (1984);
  E.~Santamato,
  Phys.\ Rev.\  D {\bf 32}, 2615 (1985).

\bibitem{Shojai:2000us}
  F.~Shojai and A.~Shojai,
  Int.\ J.\ Mod.\ Phys.\  A {\bf 15}, 1859 (2000)
  [arXiv:gr-qc/0010012];
  A.~Shojai,
  Int.\ J.\ Mod.\ Phys.\  A {\bf 15}, 1757 (2000)
  [arXiv:gr-qc/0010013];
  F.~Shojai and A.~Shojai,
  Pramana {\bf 58}, 13 (2002)
  [arXiv:gr-qc/0109052];
  F.~Shojai and A.~Shojai,
  arXiv:gr-qc/0404102;
  A.~Shojai, F.~Shojai and N.~Dadhich,
  Int.\ J.\ Mod.\ Phys.\  A {\bf 20}, 2773 (2005)
  [arXiv:gr-qc/0504137];
  F.~Shojai and A.~Shirinifard,
  Int.\ J.\ Mod.\ Phys.\  D {\bf 14}, 1333 (2005)
  [arXiv:gr-qc/0504138].

\bibitem{Carroll:2004hs}
  R.~Carroll,
  arXiv:gr-qc/0406004;
  R.~Carroll,
  arXiv:quant-ph/0406203;
  R.~Carroll,
  arXiv:math-ph/0701007;
  R.~Carroll,
  arXiv:0705.3921 [gr-qc].


\bibitem{Koch:2008hn}
  B.~Koch,
  arXiv:0801.4635 [quant-ph];
  B.~Koch,
  arXiv:0810.2786 [hep-th];
  B.~Koch,
  arXiv:0901.4106 [gr-qc];
  B.~Koch,
  arXiv:1004.2879 [hep-th].

\bibitem{Abraham:2008yr}
  J.~M.~Isidro, J.~L.~G.~Santander and P.~F.~de Cordoba,
  arXiv:0808.2351 [hep-th];
  S.~Abraham, P.~F.~de Cordoba, J.~M.~Isidro and J.~L.~G.~Santander,
  arXiv:0810.2236 [hep-th];
  S.~Abraham, P.~F.~de Cordoba, J.~M.~Isidro and J.~L.~G.~Santander,
  arXiv:0810.2356 [hep-th];
  J.~M.~Isidro, J.~L.~G.~Santander and P.~F.~de Cordoba,
  J.\ Phys.\ Conf.\ Ser.\  {\bf 174}, 012033 (2009)
  [arXiv:0902.0143 [hep-th]];
  J.~M.~Isidro, J.~L.~G.~Santander and P.~F.~de Cordoba,
  arXiv:0912.1535 [hep-th];
  J.~M.~Isidro, P.~F.~de Cordoba, J.~M.~Rivera-Rebolledo and J.~L.~G.~Santander,
  arXiv:1007.4929 [hep-th].


\bibitem{Koch:2010bz}
  B.~Koch,
  AIP Conf.\ Proc.\  {\bf 1232}, 313 (2010)
  [arXiv:1004.2879 [hep-th]];
  B.~Koch,
  AIP Conf.\ Proc.\  {\bf 1196}, 161 (2009)
  [arXiv:1004.3240 [gr-qc]].

\bibitem{Falciano:2010zz}
  F.~T.~Falciano, M.~Novello and J.~M.~Salim,
  Found.\ Phys.\  {\bf 40}, 1885 (2010).


\bibitem{CufaroPetroni:1984vw}
  N.~Cufaro Petroni, P.~Gueret, J.~P.~Vigier,
  Nuovo Cim.\  {\bf B81}, 243-259 (1984).

\bibitem{CufaroPetroni:1984wn}
 N.~Cufaro Petroni, P.~Gueret, J.~P.~Vigier,
 Phys.\ Rev.\  {\bf D30}, 495-497 (1984).

\bibitem{Holland:1988et}
  P.~R.~Holland,
  Phys.\ Lett.\  A {\bf 128}, 9 (1988);

\bibitem{Nikolic:2003yu}
  H.~Nikolic,
  Found.\ Phys.\ Lett.\  {\bf 18}, 123 (2005)
  [arXiv:quant-ph/0302152];
  H.~Nikolic,
  Found.\ Phys.\ Lett.\  {\bf 18}, 549 (2005)
  [arXiv:quant-ph/0406173].

\bibitem{Nikolic:2007ih}
 H.~Nikolic,
 Found.\ Phys.\  {\bf 39}, 1109-1138 (2009).
 [hep-th/0702060 [HEP-TH]].

\bibitem{Wetterich:2010eh}
  C.~Wetterich,
  Annals Phys.\  {\bf 325}, 2750 (2010)
  [arXiv:1006.4254 [hep-th]].


\bibitem{Holland:2008cc}
  P.~Holland,
  J.\ Phys.\ A  {\bf 42}, 075307 (2009)
  [arXiv:0807.4482 [quant-ph]].
 
\bibitem{Brown:1958zz}
  L.~M.~Brown,
  Phys.\ Rev.\  {\bf 111}, 957 (1958).

\bibitem{Feynman:1958ty}
 R.P.~Feynman and M.~Gell-Mann,
 Phys. Rev. {\bf 109}, 193 (1958);

\bibitem{CufaroPetroni:1985tu}
 N.~Cufaro Petroni, P.~Gueret, J.~P.~Vigier {\it et al.},
 Phys.\ Rev.\  {\bf D31}, 3157-3161 (1985).


\bibitem{Hill:1951zz}
  E.~L.~Hill,
  Rev.\ Mod.\ Phys.\ {\bf 23(3)}, 253 (1951).

\bibitem{Holland:2006tp}
 P.~Holland,
 Found.\ Phys.\  {\bf 36}, 369-384 (2006).




\end{thebibliography}
\end{document}